\documentclass[12pt,preprint]{aastex}

\shorttitle{VLBA imaging of the most luminous quasar}
\shortauthors{Wang et al.}

\begin{document}


\title{Milli-arcsecond imaging of the radio emission from the quasar with the most massive black 
hole at reionization}


\author{Ran Wang\altaffilmark{1},
Emmanuel, Momjian\altaffilmark{2}, 
Chris L. Carilli\altaffilmark{2},
Xue-Bing Wu\altaffilmark{1,3}, 
Xiaohui Fan\altaffilmark{4},
Fabian Walter\altaffilmark{5},
Michael A. Strauss\altaffilmark{6},
Feige Wang\altaffilmark{3},
Linhua Jiang\altaffilmark{1},
}
\altaffiltext{1}{Kavli Institute of Astronomy and Astrophysics at Peking University, 
No.5 Yiheyuan Road, Haidian District, Beijing, 100871, China}
\altaffiltext{2}{National Radio Astronomy Observatory, PO Box 0, Socorro, NM, USA 87801}
\altaffiltext{3}{Department of Astronomy, School of Physics, Peking University, No. 5 Yiheyuan Road, Haidian District, Beijing, 100871, China}
\altaffiltext{4}{Steward Observatory, University of Arizona, 933 N Cherry Ave., Tucson, AZ, 85721, USA}
\altaffiltext{5}{Max-Planck-Institute for Astronomy, K$\rm \ddot o$nigsstuhl 17, 69117 Heidelberg, Germany}
\altaffiltext{6}{Department of Astrophysical Sciences, Princeton University, Princeton, NJ, USA, 08544}

\begin{abstract}
We report Very Long Baseline Array (VLBA) observations of the 1.5 GHz radio
continuum emission of the {\it z=6.326} quasar SDSS J010013.02+280225.8
(hereafter J0100+2802). J0100+2802 is, by far the most optically luminous, and radio-quiet 
quasar with the most massive black hole known at z$>$6. The VLBA observations have 
a synthesized beam size of 12.10 mas $\times$5.36 mas (FWHM), and detected the 
radio continuum emission from this object with a peak surface brightness of $\rm 
64.6\pm9.0\,\mu Jy\,beam^{-1}$ and a total flux density of $\rm 88\pm19\,\mu Jy$. 
The position of the radio peak is consistent with
that from SDSS in the optical and Chandra in the X-ray. The radio source is marginally
resolved by the VLBA observations. A 2-D Gaussian fit to the image constrains the source
size to $\rm (7.1\pm3.5)\,mas\times(3.1\pm1.7)\,mas$. This corresponds to a
physical scale of $\rm (40\pm20)\,pc\times(18\pm10)\,pc$.
We estimate the intrinsic brightness
temperature of the VLBA source to be $\rm T_{B}=(1.6\pm1.2)\times10^{7}$ K.
This is significantly higher than the maximum value in
normal star forming galaxies, indicating an AGN
origin for the radio continuum emission. However, it is also significantly lower than the
brightness temperatures found in highest redshift radio-loud quasars.
J0100+2802 provides a unique example to study the radio activity in
optically luminous and radio quiet active galactic nuclei in the
early universe. Further observations at multiple radio frequencies
will accurately measure the spectral index and address the dominant radiation mechanism
of the radio emission.
\end{abstract}





\section{Introduction}

The large sample of quasars that have been discovered at $\rm z\geq$6 gives us the best opportunity 
to study the formation of the first supermassive black holes (SMBH) at the epoch of cosmic 
reionization (e.g., \citealp{fan06,willott10,mortlock11,jiang15,venemans15,matsuoka16,banados16}). 
Observations from X-ray to radio wavelengths indicate that the broad-band 
spectral energy distributions of most of these earliest quasars are comparable 
to those of the optically selected quasars at low-z \citep{jiang06}. 
However, their average Eddington ratio (the ratio 
between the quasar bolometric luminosity and Eddington luminosity) is close 
to unity \citep{willott10,derosa11,derosa14}.
This is higher than the average value of 0.16 found with the luminosity-matched low-z
SDSS quasar sample \citep{derosa11}, indicating that the SMBHs are 
accreting close to the Eddington limit in these young quasars at the earliest epoch. 

Radio observations of the optically selected z$>$5.5 quasars with the Very Large 
Array (VLA) at 1.4 GHz, including data from the Faint Images of the Radio Sky 
at Twenty-Centimeters (FIRST, \citealp{becker95}), yield a radio-loud fraction of $\rm 8.1^{+5.0}_{-3.2}$\%, 
which is comparable to the radio-loud fraction found with samples of low-z 
optical quasars \citep{jiang07,banados16}. Four of the radio-loud quasars at z$>$5.7 
(SDSS J0836+0054, \citealp{frey03,frey05}; SDSS J1427+3312, \citealp{frey08,momjian08};  
CFHQS J1429+5447, \citealp{frey11}; SDSS J2228+0110, \citealp{cao14}) were observed 
using the Very Long Baseline Interferometry (VLBI) with the European VLBI Network (EVN) 
and/or the Very Long Baseline Array (VLBA) at $\leq$10 
milli-arcsecond (mas) resolution at multiple frequencies. The VLBI images reveal 
compact radio emission on scales of a few tens of pc and steep radio spectra ($\rm S_{\nu}\sim\nu^{\alpha}$) 
with spectral indices of $\rm \alpha=-0.8\sim-1.1$ \citep{frey05,frey08,momjian08,frey11}. 
The EVN and VLBA images of SDSS J1427+3312 resolved the radio emission 
into two components separated by about 170 pc, indicating a very young symmetric radio structure 
with possible kinematic age of only $\rm 10^{3}$ yr \citep{frey08,momjian08}.

In this paper, we report VLBA observations of a radio-quiet quasar, SDSS J$\rm 010013.02+280225.8$
(hereafter J0100+2802) at z=6.326 \citep{wu15,wang16}. J0100+2802 is the most optically 
luminous quasar discovered at z$>$6, which hosts the most massive SMBH of 
$\rm M_{BH}\approx1.2\times10^{10}\,M_{\odot}$ among the known z$>$6 quasars \citep{wu15}. 
The X-ray to near-infrared observations yield a quasar bolometric luminosity close to the 
Eddington luminosity \citep{wu15,ai16}, suggesting that we are witnessing the rapid 
accretion of this very massive SMBH.
Our VLA observations at 3 GHz detected the radio continuum from this object with an observed 
flux density of $\rm S_{3GHz}=104.5\pm3.1\,\mu Jy$. The radio and optical 
data estimate the radio loudness of this object to 
be $R=f_{\nu,5GHz}/f_{\nu,4400\AA}=0.9$ for a steep radio spectrum ($\rm S_{\nu}\sim\nu^{-0.9}$) or R=0.2 for a flat spectrum 
($\rm S_{\nu}\sim\nu^{-0.06}$, see the discussion in \citealp{wang16}). This indicates that the central AGN 
is radio-quiet\footnote{We here adopt $R=f_{\nu,5GHz}/f_{\nu,4400\AA}=10$ to 
separate radio-loud and radio-quiet quasars \citep{kellermann89}.} but not radio-silent. J0100+2802 is the best example to study the radio activity 
from optically luminous but radio-quiet quasars at the earliest epoch. 

The VLBA observations presented in this paper provide the first 10-mas resolution ($\sim$60 pc at the quasar redshift) image 
of a radio-quiet quasar at the highest redshift, which allows us to constrain the source size and brightness temperature of the 
radio emission from this powerful AGN in the early universe.
The observation and data reduction are described in Section 2, the results are presented 
and discussed in Section 3, and the main conclusions are summarized in Section 4.
We adopt a $\Lambda$CDM cosmology with $\rm H_{0}=71km\,s^{-1}\,Mpc^{-1}$,
$\rm \Omega_{M}=0.27$ and $\rm \Omega_{\Lambda}=0.73$ throughout this paper \citep{spergel07}. 

\section{Observations and Data reduction}

We observed the 1.5 GHz radio continuum emission from J0100+2802 using the VLBA. The
observations were carried out in February and March 2016 in eight separate observing sessions.
The data at each VLBA station were recorded using the Roach Digital Backend (RDBE) and the
polyphase filterbank digital signal-processing algorithm (PFB) with sixteen 32 MHz data channels.
This provided eight contiguous 32 MHz data channels at matching frequencies in each of the two
circular polarizations. The data were sampled at two bits. We employed nodding-style phase-referencing
with a cycle time of 5 minutes: 4 minutes on the target and 1 minute on the phase calibrator, J0057+3021,
which is 2.4$^{\circ}$ away from the target. The source 3C 84 was observed as a fringe
finder and bandpass calibrator. The total observing time was 22 hours, and the total on target-source time
was 16 hours. We used the VLBA DiFX correlator (Deller, et al. 2011) in Socorro, NM,
with an integration time of 2s. Each 32 MHz data channel was further split into 128 spectral points.

We edited and calibrated the data using the Astronomical Image Processing System (AIPS; Greisen 2003) 
following standard VLBI data reduction procedures. We then performed self-calibration on the phase 
calibrator and applied the solution to the target. The continuum emission from the target was then 
imaged adopting natural weighting. The resulting synthesized beam size (FWHM) of the final image is 
12.10 mas $\times$5.36 mas, or 68.5 pc $\times$30.3 pc at the quasar redshift of z=6.326.
We also tapered the visibility data using a Gaussian function that falls to 30\% at 5 M$\lambda$ 
in both the $u$ and $v$ directions to obtain a lower resolution image and recover more flux density in 
the extended area. We list the observing parameters and the 1$\sigma$ rms noise values 
of both the full resolution and tapered images in Table 1. 

\section{Results and discussion}

We detect radio continuum emission with a single peak from the VLBA 1.5 GHz image of the 
quasar J0100+2802. The observing frequency of 1.5 GHz corresponds to a rest-frame 
frequency of 11 GHz at the quasar redshift of z=6.326 \citep{wang16}. 
We list the peak surface brightness and the peak position of the VLBA source, as well as 
other measurements in Table 1. Considering the synthesized beam, signal-to-noise ratio, 
positional uncertainty of the phase-reference calibrator, and the angular distance between 
the reference calibrator and the target, the position accuracy of this 
measurement is $\rm \sim$1 mas \citep{reid14}. 
The VLBA peak position is about 16 mas away from the peak position (RA=01h00m13.024s, 
Dec=+28d02$'$25.80$''$) of the previous VLA 3 GHz detection \citep{wang16}. 
Note that the thermal noise introduces a position error of $\rm 0.5FWHM_{beam}/SNR\sim10\,mas$ \citep{reid14} 
in the VLA measurement, and the calibrator of the VLA observations also has an position 
uncertainy\footnote{The calibrator J0119+321 is in the C Category of position accuracy according to https://science.nrao.edu/facilities/vla/observing/callist} which can range between 10 and 150 mas.  
Thus, the source position from the VLBA observations is in agreement with the VLA 3 GHz data 
within the expected errors. Figure 1 compares the position to those measured by the Sloan 
Digital Sky Survey in the optical, 
and the Chandra X-ray telescope \citep{wang16,wu15,ai16}. 
The VLBA peak is within the astrometric errors 
of the optical and X-ray observations. 

The radio emission is marginally resolved by the VLBA. 
We performed 2-D Gaussian fitting to the VLBA image using the IMFIT task in 
the Common Astronomy Software Applications package (CASA, \citealp{mcmullin07}).
The deconvolved FWHM source size (see Table 1) corresponds to a physical 
scale of $\rm (40\pm20)\,pc\times(18\pm10)\,pc$, i.e., $\rm \sim3.4\times10^{4}$ 
Schwarzschild radii, at the quasar redshift.
The total source flux density from the Gaussian fit is $\rm 88\pm19\,\mu Jy$. 
In order to better recover the total flux density of this radio component, 
we tapered the visibility data at 5 M$\lambda$. The tapered image shows an unresolved source with a peak 
flux density of 91$\pm$17 $\mu$Jy, which is consistent with the total flux 
density measured with the full resolution image. Based on the total 1.5 GHz 
flux density and the deconvolved source size from the full resolution image, 
we derived the intrinsic brightness temperature to be $\rm T_{B}=(1.6\pm1.2)\times10^{7}$ K. 
This is more than two orders of magnitude higher than the maximum brightness 
temperature in normal star-forming galaxies (i.e., $\rm T_{B}\leq10^{5}\,K$ 
at $\rm \nu\geq 1\,GHz$, \citealp{condon92}), indicating an AGN origin for the radio emission. 
However, this brightness temperature is significantly lower than the values of $\rm \sim10^{8}$ to $\rm 10^{9}$ K
found for the radio-loud quasars at z$\sim$6 at a similar frequency\citep{momjian08,frey11,cao14}. 

There are no low resolution observations of this quasar at 1.5 GHz. The flux density not included in the VLBA 
observations, if any, is unknown, and thus the spectral index of the radio continuum from 3 to 1.5 GHz (i.e., 
22 to 11 GHz in the quasar rest-frame) is unknown. 
Our previous 3 GHz and 32 GHz observations with the VLA suggest a steep power-law spectrum of $\rm S_{\nu}\sim\nu^{-0.9\pm0.15}$
from rest-frame 234 to 22 GHz, while the flux densities measured in the 
observing frequency window of 2-4 GHz (around 22 GHz in the rest-frame) 
prefer a flat spectrum of $\rm S_{\nu}\sim\nu^{-0.06\pm0.22}$ \citep{wang16}.
The 1.5 GHz flux densities, derived with the 3 GHz data of $\rm S_{3GHz}=104.5\pm3.1\,\mu Jy$ and
the steep and flat spectra described above, are $\rm 195\pm21\,\mu$Jy and $\rm 109\pm17\,\mu$Jy, respectively.

The VLBA detection of $\rm 88\pm19\,\mu Jy$ is lower, but marginally 
consistent with the flat spectral estimate. However, it is only 45\% 
of the value in the steep spectrum case. 
If there is no more diffuse emission, i.e., the VLBA source dominates the 1.5 GHz radio emission from J0100+2802, 
the VLA and VLBA data may indicate a radio spectrum that is steep at 200 to 20 GHz 
and which flattens/turns over at 20 to 10 GHz. 
However, we notice that at the rest-frame frequency of 11 GHz, the brightness temperature 
of $\rm 1.6\times10^{7}\,K$ is too low for radio emission turnover 
produced by synchrotron self absorption \citep{gallimore96}. 
Low brightness temperatures of $\rm 10^{6}$ to $\rm 10^{7}$ K and flat/turnover spectra have been  
discovered in VLBA imaging of low-z AGNs \citep{gallimore04,krips07}, 
which may be explained as free-free or synchrotron emission from an X-ray heated corona or disk 
wind \citep{blundell07,krips06,raginski16}. 
Further observations at multiple frequencies are needed to 
properly measure the radio spectral index in the 1 to 100 GHz frequency range,  
determing whether the radio spectrum of J0100+2802 is similar to the steep spectra found with 
the radio-loud objects, or if it flattens toward centimeter wavelengths.
These, together with the VLBA image will finally constrain the radiation mechanism of 
the radio emission from this extremely luminous and radio-quiet quasar at the highest redshift. 

\section{Conclusions}

We presented VLBA 1.5 GHz observations of the radio continuum emission from the 
radio-quiet quasar J0100+2802 at z=6.326. The VLBA observation reveals a compact radio 
source on scales of $\sim$40$\pm$20 pc, with a total 
flux density of $\rm 88\pm19\,\mu Jy$ (see Table 1). This is much lower than the
flux density derived from a steep power-low spectrum of $\rm S_{\nu}\sim\nu^{-0.9}$
based on previous VLA 3GHz and 32 GHz observations. 
This is the first time the radio emission  
from a radio-quiet quasar at z$>$6 has been imaged with the VLBA. 
The peak position of the VLBA source is within the 
uncertainties of the quasar locations measured by SDSS in 
the optical and Chandra in the X-ray; no clear offset 
is detected between the radio emission and the optically luminous AGN. 
We estimate the brightness temperature of the VLBA source 
to be $\rm (1.6\pm1.2)\times10^{7}$ K. This is much higher than maximum values of 
normal star-forming galaxies, indicating that radio activity from the central AGN 
is the dominant source of the VLBA detection.
J0100+2802 provides a unique example for further radio observations at multiple 
frequencies and a detailed study of the radiation mechanism of the young and 
radio-quiet quasars at the highest redshift.

\acknowledgments
The data presented in this paper are based on observations of VLBA project 16A-247.
The National Radio Astronomy Observatory
is a facility of the National Science Foundation operated under
cooperative agreement by Associated Universities, Inc.
This work made use of the Swinburne University of Technology software correlator, 
developed as part of the Australian Major National Research Facilities 
Programme and operated under licence.
We are thankful for the supports from the National
Science Foundation of China (NSFC) grants No.11373008, 11533001,
the Strategic Priority Research Program $'$The Emergence of Cosmological Structures$'$
of the Chinese Academy of Sciences, grant No. XDB09000000,
the National Key Basic Research Program of China 2014CB845700. 
This work was supported by National Key Program for Science and Technology 
Research and Development (grant 2016YFA0400703). 
RW acknowledge supports from the Thousand Youth
Talents Program of China, the NSFC grants No. 11443002 and 11473004.
XF acknowledge supports from NSF Grants AST 11-07682 and 15-15115.


{\it Facilities:} \facility{VLBA}

\clearpage


\begin{table}
\caption{VLBA observation of J0100+2802}
\begin{tabular}{lc}
\hline \noalign{\smallskip}
\hline \noalign{\smallskip}
Total observing time (hr) & 22 \\
Observing frequency (GHz) & 1.5\\
Total bandwidth (MHz) & 256 \\
Phase calibrator & J0057+3021 \\
\multicolumn{2}{l}{\em Full resolution image of J0100+2802:} \\
FWHM beam size (mas) & 12.10$\times$5.36$^{a}$ \\
Beam P.A. ($^{\circ}$) & 5 \\
Image 1$\sigma$ rms sensitivity ($\rm \mu Jy\,beam^{-1}$) & 9 \\
Peak position (J2000) & 01:00:13.0250 +28:02:25.791 \\
Peak surface brightness ($\rm \mu Jy\,beam^{-1}$) & $\rm 64.6\pm9.0$ \\
Total flux density ($\rm \mu Jy$) & $\rm 88\pm19.0$ \\
Deconvolved FWHM source size (mas) & $\rm (7.1\pm3.5)\times(3.1\pm1.7)$ \\
P.A. ($^{\circ}$) & $\rm 170\pm14$ \\
Brightness temperature (K) & $\rm (1.6\pm1.2)\times10^{7}$ \\
\multicolumn{2}{l}{\em Tapered image of J0100+2802:} \\
FWHM beam size (mas) & 34.1$\times$28.3$^{a}$  \\
Beam P.A. ($^{\circ}$) & 32 \\
Image 1$\sigma$ rms sensitivity ($\rm \mu Jy\,beam^{-1}$) & 17 \\
Total flux density ($\rm \mu Jy$) & 91$\pm$17 \\
\noalign{\smallskip} \hline
\end{tabular}\\
Note -- $^{a}$Natural weighting.
\end{table}
\begin{figure}
\epsscale{.80}
\plotone{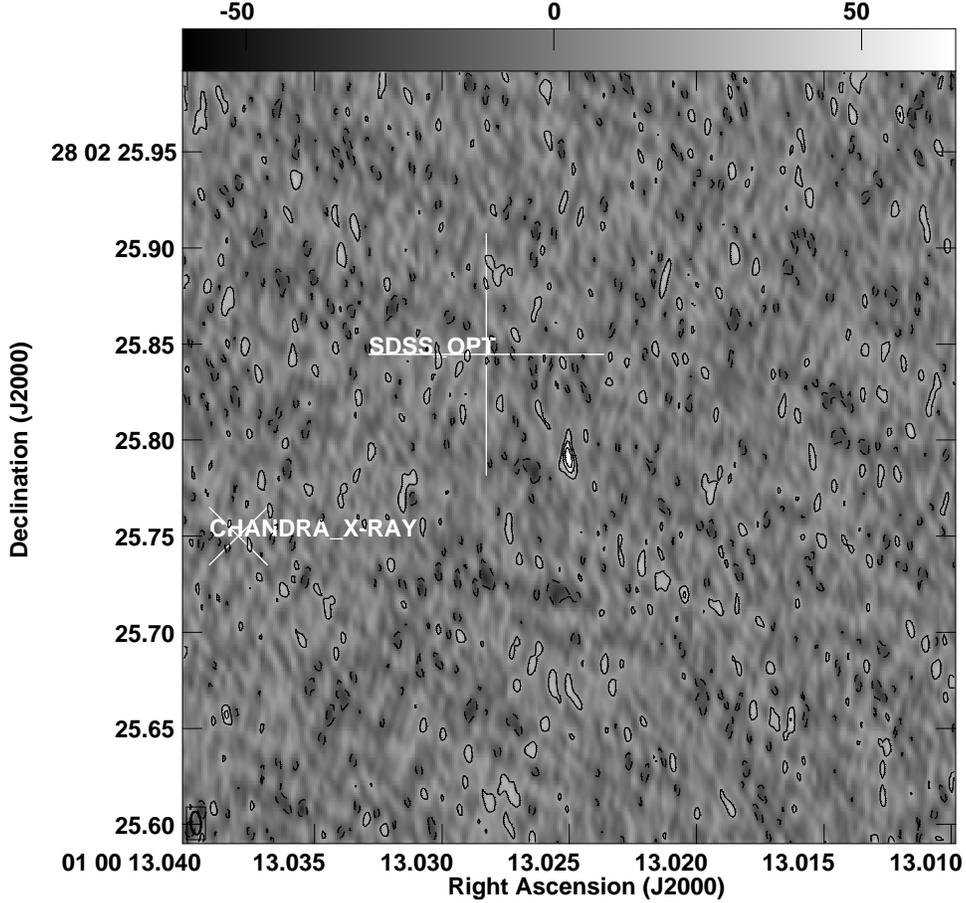}
\caption{The radio emission from J0100+2802 detected by the VLBA at 1.5 GHz, 
compared to the positions of the quasar measured from SDSS in the optical 
and Chandra in X-ray band. The image is centered on the peak position 
of the VLBA detection, and the contours are [-2, 2, 4, 6]$\rm \times9\,\mu Jy\,beam^{-1}$. 
The step-wedge at the top of the image shows the gray scale in units of $\rm \mu Jy\,beam^{-1}$.
The white plus sign represents the position and uncertainties ($\sim$0.06$''$) in RA and Dec from SDSS. 
The cross shows the X-ray source position detected by Chandra \citep{ai16}.
The Chandra observation has a position uncertainty of 0.6$''$ \citep{ai16}, 
which is larger than the field of view of $\rm 0.4''\times0.4''$ shown 
here. The synthesized beam is plotted at the bottom 
left of the image \label{fig1}
}
\end{figure}
\begin{figure}
\plottwo{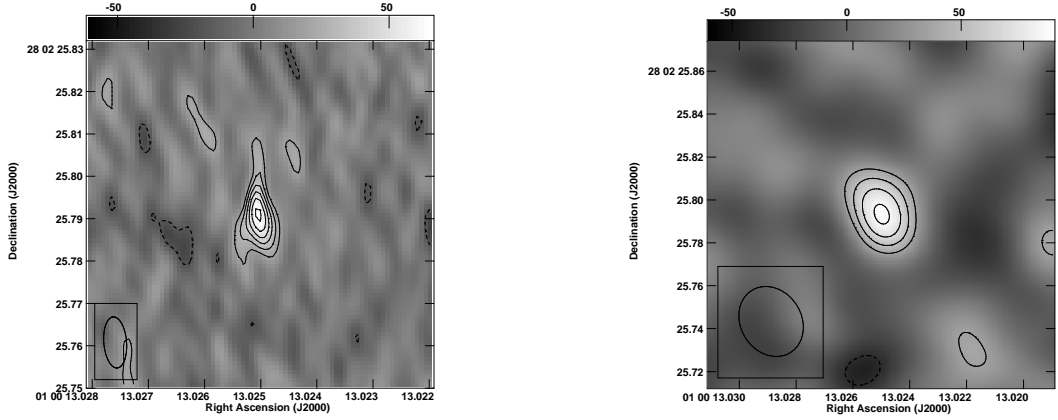}{fig2b.eps}
\caption{{\bf Left} -- The full resolution 1.5 GHz image of 
J0100+2802 (the same as Figure 1), zooming in to a smaller field of 
view of $\rm 0.08''\times0.08''$ to show the radio source detected by the VLBA. 
The contours are [-3, -2, 2, 3, 4, 5, 6, 7]$\rm \times9\,\mu Jy\,beam^{-1}$.
{\bf Right} -- Image made by the visibility data tapered at 5 M$\lambda$. 
The contours are [-2, 2, 3, 4, 5]$\rm \times17\,\mu Jy\,beam^{-1}$, 
and the synthesized beam is shown at the bottom left. 
The step-wedges at the tops of both images show the gray scale in 
units of $\rm \mu Jy\,beam^{-1}$.
The elliptics on the bottom-left of each image show the FWHM beam sizes 
of the full resolution and the tapered images (see Table 1).
Natural weighting is adopted for both images.
}
\end{figure}

\end{document}